# Structural transition and superconductivity in hydrothermally synthesized FeX (X = S, Se)

Ursula Pachmayr, Natalie Fehn, and Dirk Johrendt

*Department Chemie, Ludwig-Maximilians-Universität München, Butenandtstr. 5-13 (Haus D), 81377 München, Germany*

**Tetragonal β-FeSe obtained by hydrothermal reaction is not superconducting and transforms to a triclinic structure at 60 K unlike superconducting FeSe from solid state synthesis, which becomes orthorhombic at 90 K. In contrast, tetragonal iron sulphide FeS from hydrothermal synthesis is superconducting at 4.8 K but undergoes no structural transition. Our results suggest that the absence of superconductivity in hydrothermally synthesized FeSe may be associated to the low-temperature structure with zigzag chains of iron atoms, which is different from the known orthorhombic *Cmme* structure of superconducting FeSe.**

Unconventional superconductivity in iron-arsenides and selenides with transition temperatures up to 56 K in bulk phases[1-3] or even exciting 100 K in thin FeSe films[4] triggers enormous interest in the scientific community.[5-9] One of the most intriguing traits of these materials is that superconductivity coexists or competes with other types of electronic, magnetic, or structural orders which may or may not directly couple to superconductivity.[10,11] Most of the iron arsenides, among them LaOFeAs and BaFe$_2$As$_2$, traverse tetragonal-to-orthorhombic phase transitions accompanied by antiferromagnetic order.[12,13] Superconductivity emerges during suppression of the magnetic order by doping or pressure, and the highest critical temperatures occur in the undistorted tetragonal phases. Such a structural transition also occurs in the iron chalcogenide FeSe with tetragonal *anti*-PbO type structure,[14] but no magnetic order follows. This was initially quite surprising since magnetism was believed to be the driving force for the lattice distortion in iron arsenides (spin-nematic),[15,16] and moreover, magnetic fluctuations were considered as important for the formation of the Cooper pairs. Recent studies conclude that the structural transition in FeSe has no magnetic origin but is a consequence of orbital ordering (orbital-nematic)[16] with an unequal occupation of the iron $3d_{xz}/3d_{yz}$ orbitals.[17,18] The latest results suggest that orbital ordering and superconductivity compete in FeSe at low temperatures.[18] Thus superconducting, orbital and structural order parameters are uniquely intertwined and display the signature of unconventional superconductivity in FeSe which enables high critical temperatures. This is in line with the fact that the relatively low $T_c$ of 8 K in pure FeSe strongly increases under pressure to 36 K and by intercalation with molecular[19] or other species to 43 K.[20,21]

Recently Lai *et al*. reported that also iron sulphide FeS (*anti*-PbO-type; mackinawite) synthesized by a hydrothermal process is superconducting at 5 K.[22] So far all efforts made to pursue superconductivity in FeS from conventional synthesis failed. However, the complexity of the Fe-S phase diagram makes the synthesis of stoichiometric FeS difficult. Contrary to FeSe, several polymorphs of FeS are known,[23,24] where the mackinawite is of near FeS composition (Fe$_{1+x}$S, $0 < x < 0.07$).[25,26] Thus, one might assume that only the low-temperature hydrothermal process used by Lai *et al*. produces stoichiometric FeS which is not accessible by conventional high-temperature routes.

Given the above scenario of FeSe the question arises, whether superconductivity in FeS also occurs in an orthorhombic phase as in the selenide. This would be a strong hint to unconventional pairing, and thus for the potential of FeS to exhibit much higher critical temperatures upon intercalation or other chemical modification. The unexpected observation of superconductivity in iron sulphide motivated us to study the low temperature crystal structures of both FeSe and FeS synthesized under mild hydrothermal conditions.

Figure 1 shows the X-ray powder pattern of FeSe obtained by hydrothermal reaction method, referred to as FeSe$_{hydro}$ in the following. The Rietveld-analysis was carried out using the structural model of *anti*-PbO type FeSe. No impurity phases occur within the experimental limits (~1 % of a crystalline phase). Chemical analysis by ICP-AAS confirmed the composition Fe$_{1.02(1)}$Se and Fe$_{1.01(1)}$Se for FeSe$_{hydro}$ and conventionally synthesized FeSe (FeSe$_{conv}$), respectively. Crystallographic parameters are listed in Table 1 together with data for FeSe$_{conv}$. The lattice parameters and the selenium $z$ positions are mutually the same, thus both crystal structures are identical from the view of X-ray powder diffraction.

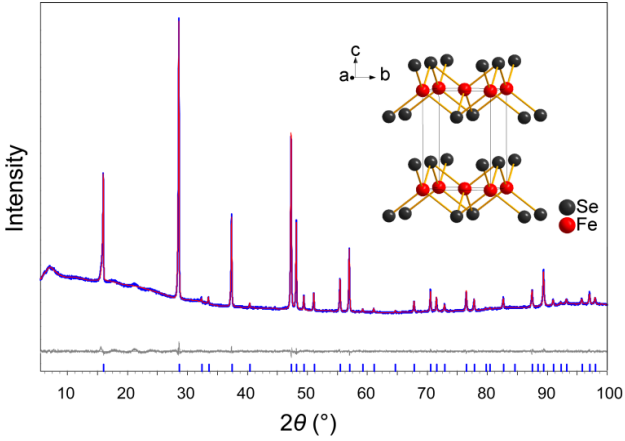

Figure 1: X-ray powder pattern of FeSe synthesized via hydrothermal reaction method (blue) with Rietveld-fit (red) and difference plot (grey). Insert: Crystal structure of *anti*-PbO type FeSe.

drothermal synthesis. Since no differences in composition or structure were detected at room temperature, next we have determined the low-temperature crystal structures.

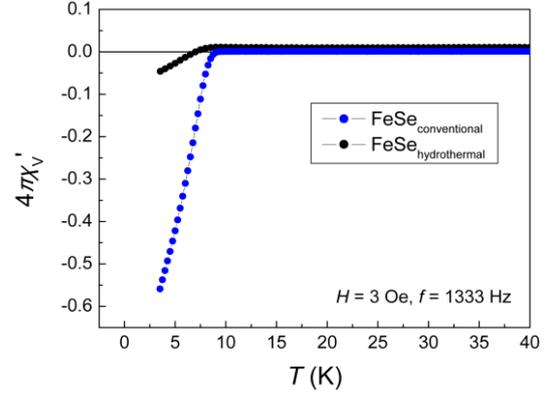

Figure 2: Low-temperature *ac*-susceptibility of FeSe samples obtained by conventional (blue) and by hydrothermal synthesis (black).

Figure 3 shows the temperature dependency of the lattice parameters. The structural transition from tetragonal (*P*4/*nmm*) to orthorhombic (*Cmme*) symmetry occurs near 90 K in FeSe$_{conv}$ in good agreement with published data.[27, 28] The transition temperature is significantly lower in the hydrothermally synthesized sample, where the lattice parameters split near 60 K.

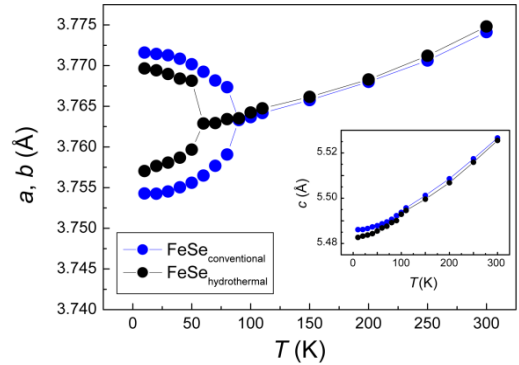

Figure 3: Temperature dependency of the lattice constants in FeSe synthesized via hydrothermal (black) and conventional (blue) reaction method, respectively. The $a$ and $b$ lattice constants are divided by $\sqrt{2}$ at temperatures below the tetragonal-to-orthorhombic phase transition.

Table 1: Crystallographic data of Fe$X$ ($X$ = S, Se)

|  | FeSe$_{hydro}$ | FeSe$_{conv}$ | FeS$_{hydro}$ |
|---|---|---|---|
| Space group | *P*4/*nmm* (No. 129, O2) | | |
| $a$ (pm) | 377.11(1) | 377.09(1) | 368.18(1) |
| $c$ (pm) | 552.14(1) | 552.16(1) | 502.97(2) |
| Volume (nm$^3$) | 0.07852(1) | 0.07852(1) | 0.06818(1) |
| Positions | 2 Fe at 2$a$ (¾,¼,0) | | |
|  | 2 Se(S) at 2$c$ (¼,¼,z) | | |
|  | $z$ = 0.2672(2) | $z$ = 0.2669(2) | $z$ = 0.262(1) |
| Phase fractions (*wt*-%) and *R*-values | | | |
| Fe$X_{(PbO-type)}$ | 100 | 93.6 | 100 |
| Fe$X_{(NiAs-type)}$ | 0 | 6.4 | 0 |
| $R_{wp}$ | 1.21 | 1.01 | 1.81 |
| $R_{exp}$ | 1.09 | 0.85 | 1.22 |
| $X^2$ | 1.11 | 1.19 | 1.49 |
| Atomic distances (pm) and angles (°) | | | |
| Fe-Fe | 266.66(1) × 4 | 266.64(1) × 4 | 260.3(1) × 4 |
| Fe-$X$ | 239.31(3) × 4 | 239.40(3) × 4 | 226.5(3) × 4 |
| $X$-Fe-$X$ | 103.93(1) × 2 | 103.97(1) × 2 | 108.8(1) × 2 |
|  | 112.31(1) × 4 | 112.29(1) × 4 | 109.8(1) × 4 |

The ac-susceptibilities of the FeSe samples are surprisingly different (Figure 2). While the expected bulk superconductivity occurs near 8 K in the conventionally synthesized sample, only traces of superconductivity are visible in the sample from hy-

A detailed inspection of the diffraction pattern reveals an asymmetric splitting of the reflections in FeSe$_{hydro}$. Figure 4 shows profiles of the (220) Bragg reflection of the tetragonal phase that splits into (400) and (040) during the phase transition. Their intensities have to be equal if the structure is orthorhombic, which is true for FeSe$_{conv}$ but not for FeSe$_{hydro}$. This means that the low-temperature structure of hydrothermally synthesized FeSe is different from the known *Cmme* structure and has lower lattice symmetry.

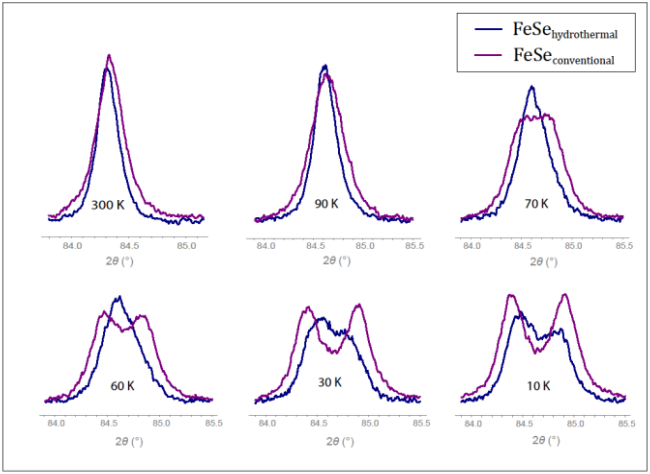

Figure 4: Temperature evolution of the $(220)_t$ Bragg reflection splitting into a doublet $[(040)_o, (400)_o]$ for FeSe synthesized via hydrothermal (blue) and conventional (magenta) reaction method, respectively.

The Rietveld refinement suggests a triclinic crystal structure ($P\bar{1}$) at 10 K with lattice parameters $a = 376.59(2)$ pm, $b = 376.66(2)$ pm and $c = 547.93(1)$ pm. The α angle remains close to 90° (90.024(4)°), β and γ alter into 89.943(4)° and 90.168(2)°, respectively. This must not be confused with an earlier assumed but never confirmed triclinic distortion in FeSe$_{0.88}$, which occurs at much higher temperature (105 K) and displays symmetric line splitting.[29] However, our low-temperature crystal structure of FeSe$_{hydro}$ differs significantly from orthorhombic superconducting FeSe$_{conv}$ and exhibits another distortion motif of the iron atoms, depicted in Figure 5. In the known orthorhombic (*Cmme*) structure, iron atoms form stripes running along the shorter axis. The four identical Fe-Fe bonds in the tetragonal phase split into two slightly shorter (265.9 pm) and two longer ones (267.2 pm),[30] however, this difference is rather tiny. In the new structure of hydrothermally synthesized non superconducting FeSe$_{hydro}$ we observe iron atoms in zigzag-chains with short Fe-Fe bonds (256.9(2), 257.7(2) pm), while the distances between neighbouring chains become long (275.2(2) pm, 276.0(2) pm). Thus the structural transition in FeSe$_{hydro}$ leads to significantly enhanced Fe-Fe bonds in the zigzag chains, while the distortion in FeSe$_{conv}$ is much weaker and the Fe-Fe bonds remain longer.

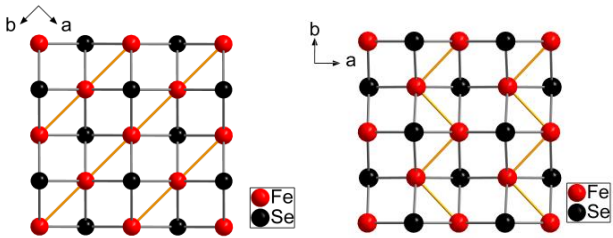

Figure 5: Low-temperature phase of FeSe synthesized via conventional (left) and hydrothermal (right) method. Iron stripes respectively iron zigzag chains are formed by short and large Fe-Fe distances.

These intriguingly different crystal structures may be the reason for the absence of superconductivity in hydrothermally prepared iron selenide. Currently it is accepted that the tiny distortion of FeSe$_{conv}$ is a result of orbital ordering, which is believed to be related to superconductivity.[18] Our results suggest that the stronger distortion in FeSe$_{hydro}$ is rather driven by Fe-Fe bond formation, which may suppresses superconductivity. However, even if the absence of superconductivity may finally be traced back to the different crystal structure, it remains unclear why the obviously identical room temperature FeSe phases transform to different low-temperature structures.

If superconductivity in FeSe only occurs in the orthorhombic phase, the question arises if this is also the case in the newly discovered superconducting FeS. We have synthesized the iron sulphide using a similar hydrothermal procedure as recently described by Lai *et al.*[22] X-ray powder diffraction revealed single-phase samples of FeS with *anti*-PbO type structure. The lattice parameters $a = 368.18(1)$ pm and $c = 502.97(2)$ pm are in good agreement with literature.[31-34] Additional X-ray single-crystal analysis confirms the tetragonal structure (see supplementary information). Our samples show superconductivity at 4.5-5 K (Figure 6). As elucidated by Lai *et al.*, the high crystallinity and high structural stability of the samples play a crucial role in the observation of superconductivity. Hydrothermal conditions turned out to be perfectly convenient to realize high quality FeS in a simple synthesis.

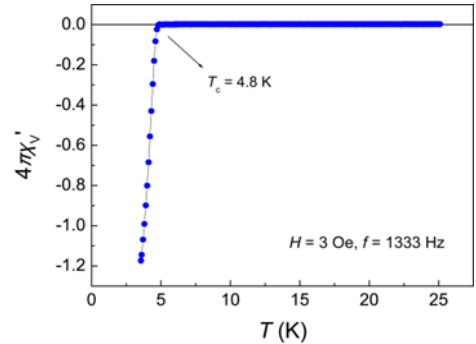

Figure 6: Low-temperature ac-susceptibility of FeS.

Figure 7 shows the temperature dependencies of the lattice parameters. As expected the unit cell decreases upon cooling, visible in a decline of the lattice parameters *a* and *c*. No anomalies are discernible and no broadening or splitting of the reflections is observed down to 10 K. Thus, contrary to FeSe, superconductivity in FeS emerges in the tetragonal phase. This scenario is reminiscent of LaOFeP and LaOFeAs. While the phosphide is a conventional superconductor with $T_c$ near 4 K, the arsenide is a parent compound of high-$T_c$ materials and exhibits magnetic fluctuations as well as a tetragonal-to-orthorhombic structural distortion. The absence of the latter in FeS indicates that the iron sulphide may rather be a conventional BCS-type superconductor and thus quite different from the selenide FeSe.

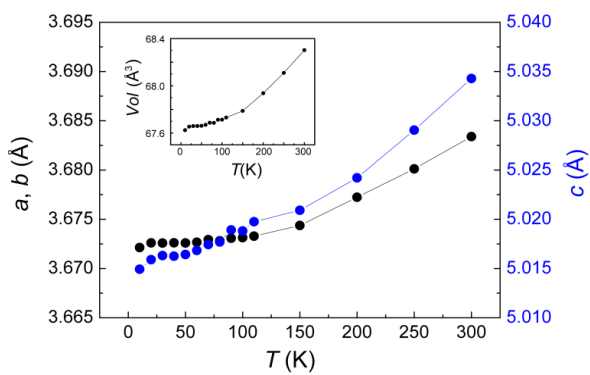

Figure 7: Lattice parameters and unit cell volume (insert) of tetragonal FeS.

Finally it remains intriguing that hydrothermal synthesis under mild conditions yields superconducting FeS but non superconducting FeSe, while the opposite is true for high-temperature solid state methods. While truly stoichiometric FeS is probably only accessible by the hydrothermal method due to the complex phase diagram, we currently have no explanation for the surprising differences of the structures and properties between the FeSe samples.

We thank for financial support by the DFG (German Research Foundation), project JO257/7-1.

## Notes and references

*E-Mail: johrendt@lmu.de*

Materials: Fe powder (Chempur, 99.9 %), Se powder (Chempur, 99.999 %), $SCN_2H_4$ crystals (Grüssing, 99 %), $NaBH_4$ powder (Acros, 98 %), NaOH pellets (Grüssing), KOH platelets (AppliChem).
Hydrothermal synthesis of Fe$X$ ($X$ = Se, S) was carried out using 1 mmol iron metal and selenium respectively thiourea as starting materials. For the synthesis of FeSe, 110 mg $NaBH_4$ was added as reducing agent and KOH as mineralizer. FeS was synthesized using NaOH as mineralizer and only 5 mg $NaBH_4$. The educts were mixed with distilled water (20 respectively 5 mL), sealed in Teflon-lined steel autoclaves (50 mL) under argon atmosphere and heated at 150 °C for 8-13 days. The black precipitates were collected by centrifugation and washed with distilled water and ethanol. During this washing step small amounts of unreacted Fe can be removed with a magnet. The samples were dried at room temperature under dynamic vacuum and stored in a purified argon atmosphere glove box. For conventional solid-state reaction method stoichiometric amounts of Fe and Se were heated under argon atmosphere for 48 h at 700 °C and 10 days at 320 °C. Powder X-ray diffraction was carried out using a Huber G670 diffractometer (Ge-111 monochromator; Cu-K$\alpha_1$ radiation; $\lambda$ = 154.05 pm) at room temperature. For low temperature, Co-K$\alpha_1$ radiation ($\lambda$ = 179.02 pm) and a close-cycle He-cryostat was used. Structural parameters were obtained by Rietveld refinement using the TOPAS package.[35] Single-crystal analysis was performed on a Bruker D8-Quest diffractometer (Mo-K$\alpha_1$, $\lambda$ = 71.069 pm, graphite monochromator). The structure was refined with the Jana2006 program package.[36] Superconductivity was examined in ac-susceptibility measurements.